\documentclass[conference]{IEEEtran}
\usepackage{graphicx,cite,epsfig,amssymb,amsmath}
\usepackage{pifont}
\usepackage{stmaryrd}
\usepackage{bbding}
\usepackage{subfigure}
\usepackage{algorithm}
\usepackage{algorithmic}
\usepackage{mathrsfs}
\usepackage{latexsym}
\usepackage{indentfirst}
\usepackage{fmtcount}
\usepackage{cite}

\usepackage{url}

\usepackage{bigstrut,multirow,rotating}

\IEEEoverridecommandlockouts

\begin{document}

\title{Enhancing User Experience for Multi-Screen Social TV Streaming over Wireless Networks}

\author{\authorblockN{Huazi~Zhang, Yichao~Jin, Weiwen~Zhang, and Yonggang~Wen}
\\School of Computer Engineering, Nanyang Technological University, Singapore.
\\Email: \{zhanghz,yjin3,wzhang9,ygwen\}@ntu.edu.sg
}

%
\maketitle


\begin{abstract}
Recently, multi-screen cloud social TV is invented to transform TV into social experience. People watching the same content on social TV may come from different locations, while freely interact with each other through text, image, audio and video. This crucial \emph{virtual living-room experience} adds social aspects into existing performance metrics. In this paper, we parse social TV user experience into three elements (i.e., inter-user delay, video quality of experience (QoE), and resource efficiency), and provide a joint analytical framework to enhance user experience. Specifically, we propose a cloud-based optimal playback rate allocation scheme to maximize the overall QoE while upper bounding inter-user delay. Experiment results show that our algorithm achieves near-optimal tradeoff between inter-user delay and video quality, and demonstrates resilient performance even under very fast wireless channel fading.
\end{abstract}


\section{Introduction}
Cloud social TV is transforming the traditional TV watching habit into a social networking experience. As users' media consuming behaviors are migrating from TV screens to smartphones and tablets \cite{yonggang-TMM-2014}, multi-screen cloud social TV \cite{montpetit2012social,yu2010hybrid,cesar2008usages,jin2013multi} was proposed to re-invent traditional TV experience through interactive social features. Specifically, its \emph{virtual living-room} experience brings interactivity among peer viewers \cite{li2013presentation}, such that various forms of information (i.e., text, image, audio and video) are exchanged in response to the ongoing video content.

However, delivering good user experience over heterogenous networks poses tremendous challenge in the real deployment of cloud social TV. \emph{First}, due to the inherent nature of stochastic wireless network, the video quality deteriorates under deep fading wireless channels. \emph{Second}, the inter-user delay accumulates as the video plays, which can affect the social interaction experience. Imagine that Peter and his friend Cathy are in a social TV ``room" watching the same soccer game. Peter excitedly texts to Cathy about a goal he just witnessed, which turns out to be ``spoiler" to Cathy who is suffering from long delay. \emph{Moreover}, the quality of service should be guaranteed within affordable operational cost, i.e., without incurring high operational cost of cloud resource. Therefore, an effective design of cloud social TV should be in place to enhance user experience by simultaneously addressing the above three issues.


Existing works in this domain separately considered either the enhancement of video quality of experience (QoE) or the reduction of operation cost of multimedia services, but none of them jointly consider the three-element user experience of cloud social TV. In \cite{Rejaie1999,li2014streaming}, adaptive video playback rate at the end users is introduced to optimize perceived video quality under rapidly changing bandwidth. Both \cite{shen2013information} and \cite{Weiwen-TMM-2013} considered the QoE-driven cache management at the server, while \cite{joseph2013nova} considered optimizing user QoE over network. But they did not include the inter-user delay for user interaction, and thus their approach cannot be adopted for social TV. On a different track, resource efficient video delivery has been studied in \cite{llorca2012energy}, in which a near-optimal energy tradeoff between video transport and processing is obtained. In \cite{jin2013toward}, the authors aimed to balance the tradeoff between content transmission cost and storage cost. But they only considered the operation cost, without the consideration of user experience.

In this paper, leveraging the real cloud social TV system \cite{jin2013multi,li2013presentation} we implemented over a cloud-centric media platform \cite{jin2012codaas}, we jointly manage video QoE, inter-user delay and cost efficiency. Specifically, we maximize the video QoE while satisfying the operation cost and the inter-user delay constraints, by optimally setting the playback rate for users. We formulate a constrained optimization problem and obtain the optimal solution of the rate allocation for two cases (i.e., moderate delay case and severe delay case), respectively. We then implement a rate allocation scheme to enhance the user experience in cloud social TV. Experiment results show that our proposed scheme outperforms two alternative strategies (i.e., maximal QoE scheme and minimal delay scheme). Our solution sheds light on the design of future commercial social TV systems with good user experience.

Our contributions are multifold:
\begin{itemize}
  \item First, we provide a joint optimization framework to enhance the overall user experience for all the users. To the best of our knowledge, this is the first work to consider \emph{inter-user delay} from the viewpoint of social interaction experience.
  \item Second, we solve the optimization problem and propose a playback rate allocation scheme to enhance the overall QoE within tolerable inter-user delay and constrained resource consumption. In addition, we provide implementation details of the scheme in our social TV platform.
  \item Finally, we verify the effectiveness of the proposed scheme through experiments, and investigate those practical issues encountered (e.g., rate adaptation, discrete video quality) during implementation. The results show that our scheme achieves near-optimal tradeoff between inter-user delay and video quality, and demonstrates resilient performance even under fast channel fading.
\end{itemize}

The rest of the paper is organized as follows. Section \ref{sect-model} presents the system model and problem formulation. We solve the optimization problem in Section \ref{sect-solution}. Section \ref{sect-result} presents the numerical analysis and results. We conclude this paper in Section \ref{sect-conclusion}.

\section{System Model and Problem Formulation}\label{sect-model}
\subsection{System Architecture}
\begin{figure}[h] \centering
\includegraphics[width=0.45\textwidth]{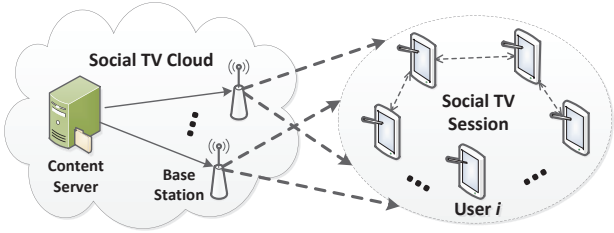}
\caption {Multi-Screen Cloud Social TV Systems Overview}
\label{system-model}
\end{figure}

Our social TV architecture is shown in Fig.~\ref{system-model}. Similar to most existing cloud media systems, it provides on-demand video services through the building blocks such as content servers, wireline/wireless networks and embedded media processing and scheduling modules.

\begin{figure}[h] \centering
\includegraphics[width=0.48\textwidth]{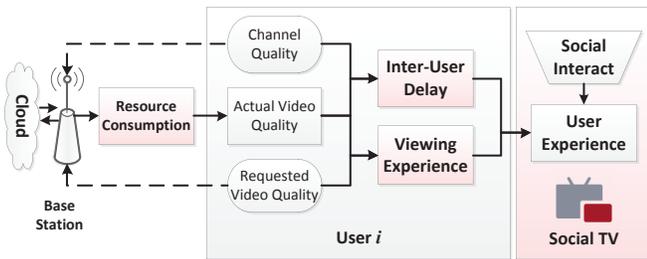}
\caption {Social TV User Experience over Wireless Networks}
\label{system-logic}
\end{figure}

The elements of social TV user experience (i.e., video experience, inter-user delay and resource consumption) and their decisive factors (i.e., the actual and requested playback rate, and the channel dynamics) are plotted in Fig.~\ref{system-logic}. When deployed over wireless networks, a primary bottleneck is the uncertainty of wireless channel quality, which influences user experience in terms of both inter-user delay and video quality. Meanwhile, the requested video quality, together with the actually obtained video quality, determines the QoE of each user. Therefore, The social TV cloud should maximize the overall perceived video quality by optimally allocating resources to users according to their dynamic channel states and requested video quality.

The major symbols used in this paper are defined as follows:
\begin{itemize}
\item $C$ - Average resource budget per user
\item $G$ - Maximum inter-user delay among users
\item $T$ - Playback duration of a video segment
\item $R_i$ - Available transmission rate of the $i$th user
\item $S_i$ - Requested segment size of the $i$th user
\item $D_i$ - Transmission delay of the $i$th user
\item $Q_i$ - Viewing experience of the $i$th user
\item ${\hat S}_i$ - Allocated segment size of the $i$th user
\end{itemize}

\subsection{Resource Constraints}
\subsubsection{Storage Model}
Upon the establishment of each session, the Social TV cloud will allocate resource for each user. Specifically, the edge server will store the video segments for each user and the base station is responsible for the transmissions of those segments. Both tasks incur certain monetary cost \cite{Yichao-GC-2013}. For each resolution corresponding to the actual segment size of user $i$, the storage cost is
\begin{equation*}
C_{storage}^i = a{\hat S_i},
\end{equation*}
where the storage cost is $a$ unit per bit.
\subsubsection{Wireless Communication Model}
We consider a dynamic channel model in which the available data rate may change over time, depending on the channel quality. Under a general wireless channel $y = \sqrt{g_i} x + n$, where $g_i$ is the power gain and $n$ is additive white Gaussian noise, the $i$th user's available transmission rate is given by
\begin{equation}\label{rate-i}
{R_i} = W \log \left( {1 + \frac {g_i P} {N_0}} \right),
\end{equation}
where $P$ is the fixed transmission power at the BS, $W$ is the channel bandwidth, and $N_0$ is the power spectral density of noise.

Since Social TV subscribers are mostly located in urban environment with access to 3/4G services or Wi-Fi, the power gain $g_i$ is assumed as a Rayleigh distributed random variable. Denote by $h_i \triangleq \frac {g_i} {N_0}$ the channel gain-to-noise ratio, $c_0$ the cost per Watt, the wireless communication cost per bit $b_i$ becomes
\begin{equation}\label{b-i}
{b_i} = \frac{c_0 P}{{\log \left( {1 + {h_i}P} \right)}},
\end{equation}
which is solely influenced by the dynamic channel gain $h_i$.

Hence, the communication cost model is
\begin{equation*}
C_{communication}^i = {b_i}{\hat S_i}.
\end{equation*}

Finally, consider an $n$-user social TV session with total budget $C \times n$, the resource constraint reads as
\begin{equation}
C1:\quad \sum\limits_{i = 1}^n {\left( {a + {b_i}} \right){{\hat S}_i}}  \le C \times n.
\end{equation}

\subsection{Inter-User Delay Model}
In this section, we focus on the delay of each segment. For each segment, the transmission delay is mainly determined by two factors -- the video quality of the segment, and the channel quality. For the $i$th user, the former corresponds to the size of the transmitted segment ${\hat S}_i$, and the latter is measured by the available transmission rate $R_i$. By plugging in \eqref{rate-i}, we obtain the segment transmission delay as
\begin{equation}
{D_i} = \frac{{{{\hat S}_i}}}{{{R_i}}} = \frac {{\hat S}_i} {W {\log \left( {1 + {h_i}P} \right)}}.
\end{equation}

In practice, while the current video segment is being played, the next segment is being downloaded in the background at the same time. In such cases, the user will not notice any delay if the transmission delay $D_i$ is shorter than the segment play time $T$. For this reason, we define a more meaningful ``inter-user play delay" as follows
\begin{equation}
D_i^p = {\left( {{D_i} - T} \right)^ + } = \left\{ {\begin{array}{*{20}{c}}
{0,}&{{D_i} \le T,}\\
{{D_i} - T,}&{{D_i} > T.}
\end{array}} \right.
\end{equation}

To ensure the \emph{virtual living-room experience}, we introduce the following ``inter-user delay constraint" to upper bound the maximal inter-user delay among all users
\begin{equation}\label{constraint-C2}
C2:\quad \left| {D_i^p - D_j^p} \right| \le G,\quad \forall i,j \in \left\{ {1, \cdots ,n} \right\} \ \& \ i\neq j,
\end{equation}
where $G$ is the maximum tolerable inter-user delay.

\subsection{Quality of Experience Model}
The video viewing experience is a function of the requested and acquired video quality, measured by a metric called QoE. A logarithmic QoE function is initially proposed in \cite{reichl2010logarithmic} to fit the mean opinion score (MOS) from a large audience watching the same video. Due to its demonstrated accuracy especially in telecommunication services such as audio \cite{reichl2013logarithmic} and video streaming \cite{Weiwen-TMM-2013}, we adopt this logarithmic model. A practical QoE model has the following form
\begin{equation}\label{Q-i}
{Q_i} = {\alpha _1}\log \left( {{\alpha _2}\frac{{{{\hat S}_i}}}{{{S_i}}}} + \alpha_3 \right),
\end{equation}
where $S_i$ and ${\hat S}_i$ are requested and allocated segment size, respectively, and $\alpha _1$, $\alpha _2$, and $\alpha _3$ are parameters determined by the video content and other system features. They are obtained by fitting the logarithmic function to empirical data.

\subsection{Optimization Problem}
The objective of social TV service is to maximize the overall QoE for each session, by optimally allocating playback rate to every participant. To achieve this target, we need to deal with two kinds of dynamics. On the one hand, the users may request for different qualities of the video according to various aspects, e.g., a tablet usually requires higher resolution than a smart phone. Likewise, an action movie may consume higher playback rate than a scenery video. On the other hand, the wireless channel states varies greatly depending on the user's location and moving speed, affecting both the available transmission rate and the per bit cost. This in turn pose a limit on the actual playback rate provided to each user.

Mathematically, the stated problem converts to the following constrained optimization problem
\begin{equation}
{\mathop {\max }\limits_{{\bf{\hat S}}} } \quad {Q\triangleq\sum\limits_{i = 1}^n {{Q_i}} ,} \quad {s.t.} \quad {C1, \ C2.}
\end{equation}
where ${\bf{\hat S}} = \left( {{{\hat S}_1},{{\hat S}_2}, \cdots ,{{\hat S}_n}} \right)$ is the allocated segment size vector to be optimized, and $C1$ and $C2$ stand for the resource constraint and the inter-user delay constraint, respectively. We further assume that the allocated playback rate should not exceed the requested playback rate. Therefore, the optimization is formally given as
\begin{align}
\mathop {\max }\limits_{{\bf{\hat S}}} \quad &Q = \sum\limits_{i = 1}^n {{\alpha _1}\log \left( {{\alpha _2}\frac{{{{\hat S}_i}}}{{{S_i}}} + {\alpha _3}} \right)}, \\
s.t. \quad &C1:\quad \sum\limits_{i = 1}^n {\left( {a + {b_i}} \right){{\hat S}_i}}  \le C \times n,\\
&C2:\quad \left| {D_i^p - D_j^p} \right| \le G,\forall i,j \in \left\{ {1, \cdots ,n} \right\},\\
&C3:\quad 0 \le {{\hat S}_i} \le {S_i},\forall i \in \left\{ {1, \cdots ,n} \right\}.
\end{align}

\section{Optimal Playback Rate Allocation}\label{sect-solution}
\subsection{Derivation of Solution}
Through examining the optimization problem and its constraints, we have the following observations. First, the objective function is convex because its Hessian matrix is positive definite. Second, both the resource constraints $C1$ and $C3$ are linear and convex sets. However, the inter-user delay constraint $C2$ is complex and need further inspection.

Therefore, we adopt a divide-and-conquer strategy based on $C2$, which may be re-written as follows
\begin{equation*}
C'2:\quad \mathop {\max }\limits_i \left\{ {{{\left( {{D_i} - T} \right)}^ + }} \right\} - \mathop {\min }\limits_i \left\{ {{{\left( {{D_i} - T} \right)}^ + }} \right\} \le G.
\end{equation*}

It can be found that if $\mathop {\min }\limits_i \left\{ {{D_i}} \right\} \le T$, the nonlinear operator ${\left(  \cdot  \right)^ + }$ may be removed. Note that we assume $\mathop {\max }\limits_i \left\{ {{D_i}} \right\} \ge T$ to avoid trivial problem, i.e., all-zero play delay case. Thus, the original problem may be divided into two subcases, $\mathop {\min }\limits_i \left\{ {{D_i}} \right\} \le T$ and $\mathop {\min }\limits_i \left\{ {{D_i}} \right\} \ge T$.

\subsubsection{Moderate Delay Case} The former subcase corresponds to not-too-bad channel quality, and at least one user finishes transmission within $T$. In such case, since $\mathop {\min }\limits_i \left\{ {{{\left( {{D_i} - T} \right)}^ + }} \right\}=0$ by definition, $C'2$ can be replaced by a set of linear constraints as follows
\begin{equation}\label{c3a}
C2.a:\quad \left( {\frac{{{{\hat S}_i}}}{{{R_i}}} - T} \right) - 0 \le G,\forall i \in \left\{ {1, \cdots ,n} \right\}.
\end{equation}

Note that the above constraint is a linear and convex set, we use Lagrange multiplier method to solve the problem. After converting the maximization problem into the corresponding minimization problem and introducing a set of slack variables, the Lagrangian function under $C2.a$ is given by
%
\begin{align}
L\left( {{{\hat S}_i},\lambda ,{\mu _i},{\nu _i},x,{y_i},{z_i}} \right) = & - \sum\limits_{i = 1}^n {{\alpha _1}\log \left( {{\alpha _2}\frac{{{{\hat S}_i}}}{{{S_i}}} + {\alpha _3}} \right)}\nonumber\\
& + \lambda \left[ {\sum\limits_{i = 1}^n {\left( {a + {b_i}} \right){{\hat S}_i}}  + {x^2} - C \cdot n} \right]\nonumber\\
& + \sum\limits_{i = 1}^n {{\mu _i}\left( {\frac{{{{\hat S}_i}}}{{{R_i}}} + y_i^2 - T - G} \right)}\nonumber\\
& + \sum\limits_{i = 1}^n {{\nu _i}\left( {{{\hat S}_i} + z_i^2 - {S_i}} \right)}.
\end{align}

Using KKT conditions, we can find a global optimum by solving the following equations
\begin{align}
\frac{{\partial L}}{{\partial {{\hat S}_i}}} &=  - \frac{{{\alpha _1}{\alpha _2}}}{{{\alpha _2}{{\hat S}_i} + {\alpha _3}{S_i}}} + \lambda \left( {a + {b_i}} \right) + \frac{{{\mu _i}}}{{{R_i}}} + {\nu _i} = 0 \label{equationset-first} \\
\frac{{\partial L}}{{\partial \lambda }} &= \sum\limits_{i = 1}^n {\left( {a + {b_i}} \right){{\hat S}_i}}  + {x^2} - C \cdot n = 0\\
\frac{{\partial L}}{{\partial {\mu _i}}} &= \frac{{{{\hat S}_i}}}{{{R_i}}} + y_i^2 - T - G = 0\\
\frac{{\partial L}}{{\partial {v_i}}} &= {{\hat S}_i} + z_i^2 - {S_i} = 0\\
\frac{{\partial L}}{{\partial x}} &= 2\lambda x = 0\\
\frac{{\partial L}}{{\partial {y_i}}} &= 2{\mu _i}{y_i} = 0\\
\frac{{\partial L}}{{\partial {z_i}}} &= 2{\nu _i}{z_i} = 0.\label{equationset-last}
\end{align}

Since the above system of equations are not only linear but also full rank, it can be easily solved. In this paper, we directly use the solver (\emph{fsolve}) provided by Matlab.

\subsubsection{Severe Delay Case} When all users in a session suffer from deep channel fading, all transmission delays are larger than $T$. Again, the nonlinear operator ${\left(  \cdot  \right)^ + }$ can be removed, and we have
\begin{align}
&\mathop {\max }\limits_i \left\{ {{{\left( {{D_i} - T} \right)}^ + }} \right\} - \mathop {\min }\limits_i \left\{ {{{\left( {{D_i} - T} \right)}^ + }} \right\}\nonumber\\
= &\mathop {\max }\limits_i \left\{ {{D_i} - T} \right\} - \mathop {\min }\limits_i \left\{ {{D_i} - T} \right\}\nonumber\\
= &\mathop {\max }\limits_i \left\{ {{D_i}} \right\} - \mathop {\min }\limits_i \left\{ {{D_i}} \right\}.
\end{align}

Thus, $C'2$ may be replaced by a set of nonlinear constraints as
\begin{equation}\label{c3b}
C2.b:\quad \frac{{{{\hat S}_i}}}{{{R_i}}} - \mathop {\min }\limits_j \left\{ {\frac{{{{\hat S}_j}}}{{{R_j}}}} \right\} \le G,\forall i \in \left\{ {1, \cdots ,n} \right\}.
\end{equation}

Note that compared with \eqref{c3a}, $T$ is replaced by an unknown variable $\mathop {\min }\limits_j \left\{ {\frac{{{{\hat S}_j}}}{{{R_j}}}} \right\}$. Thus, we cannot directly apply the previous method. Instead, we resort to Lagrange dual method which iteratively converge to the optimum. In particular, we denote
\begin{equation}
k = \arg \mathop {\min }\limits_i \left\{ {\frac{{{{\hat S}_i}}}{{{R_i}}}} \right\}.
\end{equation}

Thus, the Lagrange function is
\begin{align}
L\left( {{{\hat S}_i},\lambda ,{\mu _i},{\theta _i},{\beta _i}} \right) =& \sum\limits_{i = 1}^n {{\alpha _1}\log \left( {{\alpha _2}\frac{{{{\hat S}_i}}}{{{S_i}}} + {\alpha _3}} \right)}\nonumber\\
& - \lambda \left[ {\sum\limits_{i = 1}^n {\left( {a + {b_i}} \right){{\hat S}_i}}  - C \times n} \right]\nonumber\\
& - \sum\limits_{i = 1}^n {\mu _i}\left( {{{\hat S}_i} - {S_i}} \right) + \sum\limits_{i = 1}^n {{\theta _i}} {{\hat S}_i}\nonumber\\
& - \sum\limits_{i = 1}^n {{\beta _i}} \left( {\frac{{{{\hat S}_i}}}{{{R_i}}} - \frac{{{{\hat S}_k}}}{{{R_k}}} - G }\right).
\end{align}

Take the derivative of ${{\hat S}_i}$, we have
\begin{equation*}
\frac{{\partial {L_i}}}{{\partial {{\hat S}_i}}} = {\alpha _1}\frac{{\frac{{{\alpha _2}}}{{{S_i}}}}}{{{a_2}\frac{{{{\hat S}_i}}}{{{S_i}}} + {\alpha _3}}} - \lambda \left( {a + {b_i}} \right) - {\mu _i} + {\theta _i} - \frac{{{\beta _i}}}{{{R_i}}} = 0.
\end{equation*}

The optimal playback rate ${\hat S_i}^*$ has the following form
\begin{equation}\label{c3b-solution}
{\hat S_i}^* = \frac{{{\alpha _1}}}{{\lambda \left( {a + {b_i}} \right) + {\mu _i} + \frac{{{\beta _i}}}{{{R_i}}} - {\theta _i}}} - \frac{{{\alpha _3}{S_i}}}{{{\alpha _2}}}.
\end{equation}

The Lagrange multipliers in \eqref{c3b-solution} are obtained in an iterative fashion using the following equations
\begin{align}
{\lambda ^{l + 1}} &= {\left[ {{\lambda ^l} - \kappa \left( {C \times n - \sum\limits_{i = 1}^n {\left( {a + {b_i}} \right){{\hat S}_i}} } \right)} \right]^ + },\label{iterations-first}\\
\mu _i^{l + 1} &= {\left[ {\mu _i^l - \kappa \left( {{S_i} - {{\hat S}_i}} \right)} \right]^ + },\\
\theta _i^{l + 1} &= {\left[ {\theta _i^l - \kappa {{\hat S}_i}} \right]^ + },
\end{align}
and most importantly, the iterative algorithm finds the minimal transmission delay, i.e., $\mathop {\min }\limits_i \left\{ {\frac{{{{\hat S}_i}}}{{{R_i}}}} \right\}$ in each round and uses it to update the following coefficient
\begin{align}\label{iterations-last}
\beta _i^{l + 1} = {\left[ {\beta _i^l - \kappa \left( {G + \mathop {\min }\limits_i \left\{ {\frac{{{{\hat S}_i}}}{{{R_i}}}} \right\} - \frac{{{{\hat S}_i}}}{{{R_i}}}} \right)} \right]^ + },
\end{align}
where $\kappa$ is the step size used to control the convergence speed and accuracy.

\subsection{Implementation of Optimal Solution}\label{sect-implementation}
\begin{figure}[h] \centering
\includegraphics[width=0.45\textwidth]{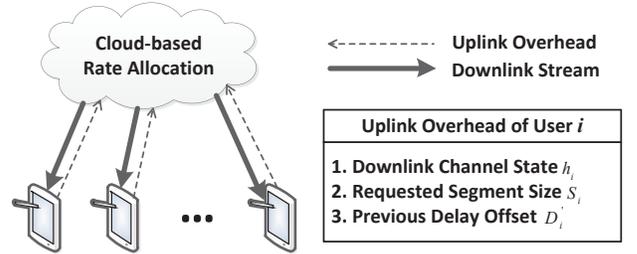}
\caption {Implementation of Cloud-based Playback Rate Allocation}
\label{Implementation}
\end{figure}

The playback rate allocation protocol is illustrated in Fig. \ref{Implementation}. At the beginning of each segment cycle, each user reports to the cloud through the uplink overhead. The latter contains important local information such as downlink channel state $h_i$, requested segment size $S_i$ and delay information of the previous segment $D'_i$. Note that the channel state $h_i$ can be easily obtained through downlink channel estimation. Altogether they serve as the inputs of the rate allocation algorithms implemented in the cloud, or more specifically, the edge server of the video source.

The social TV cloud will execute \textbf{Algorithm 1} by default and switch to \textbf{Algorithm 2} under severe delay. If history delay information is available, the server may assume that delay status remains the same and choose algorithm accordingly. Note that this is computation-efficient because the delay status, along with the channel status and user request, usually does not change very frequently. However, once a delay status change is detected, the server can freely switch between algorithms as described in the pseudo codes.

\begin{algorithm}
\caption{Moderate Delay Case} \label{Moderate}
\begin{algorithmic}
\IF {$\mathop {\min }\limits_i \left\{ {{D'_i} - T} \right\} > 0$}
\STATE Set Flag = Severe, goto \textbf{Algorithm 2}
\ELSE \FOR{$i \in 1, \cdots ,n$}
\STATE Estimate $R_i,b_i$ using $h_i, P, W$ and \eqref{rate-i}, \eqref{b-i}
\ENDFOR
\STATE Establish equation set \eqref{equationset-first}--\eqref{equationset-last}
\STATE Obtain ${\hat S}_i^*$ for all $i \in 1, \cdots ,n$ using \emph{fsolve($\cdot$)}
\IF {$\mathop {\min }\limits_i \left\{ {\frac{{{{\hat S}_i^*}}}{{{R_i}}} - T} \right\} < 0$}
\STATE Allocate ${\hat S}_i^*$ for all $i \in 1, \cdots ,n$ to each user
\ENDIF
\ENDIF
\end{algorithmic}
\end{algorithm}

\begin{algorithm}
\caption{Severe Delay Case} \label{Severe}
\begin{algorithmic}
\FOR{$i \in 1, \cdots ,n$}
\STATE Estimate $R_i,b_i$ using $h_i, P, W$ and \eqref{rate-i}, \eqref{b-i}
\STATE Assign initial values to ${\hat S}_i^0$
\ENDFOR
\WHILE {$\left| {{{{\bf{\hat S}}}^{l}} - {{{\bf{\hat S}}}^{l-1}}} \right| > \varepsilon $}
\STATE Determine $\mathop {\min }\limits_i \left\{ {\frac{{{{\hat S}_i}}}{{{R_i}}}} \right\}$
\STATE Update $\left\{ {{\lambda ^l},{\mu_i^l},{\theta_i^l},{\beta_i^l}} \right\}$ using \eqref{iterations-first}--\eqref{iterations-last}
\STATE Calculate ${{{\hat S}_i}}$ for all $i \in 1, \cdots ,n$ using \eqref{c3b-solution}
\STATE $l$++
\ENDWHILE
\STATE Assign ${\hat S_i^*} = {\hat S_i}$ for all $i \in 1, \cdots ,n$
\IF {$\mathop {\min }\limits_i \left\{ {\frac{{{{\hat S}_i^*}}}{{{R_i}}} - T} \right\} \ge 0$}
\STATE Allocate ${\hat S}_i^*$ for all $i \in 1, \cdots ,n$ to each user
\ENDIF
\end{algorithmic}
\end{algorithm}

Note that although our algorithm is performed within one segment cycle, it can be easily modified to run over consecutive segments. Recall the inter-user delay constraint \eqref{constraint-C2}, we only need to modify this constraint by adding the delay offset of the previous segment into the current delay as follows
\begin{align}\label{adaptive-constraint}
C''2: \quad | {{\left( {{D'_i} + {D_i} - T} \right)}^x+ } - & {{\left( {{D'_j} + {D_j} - T} \right)}^+ } | \le G,& \nonumber\\
&\forall i,j \in \left\{ {1, \cdots ,n} \right\}.
\end{align}

With the modified ``accumulative delay constraint", the residual delays of previous segments are passed on to the current segment, so that our proposed scheme can perform ``adaptive rate allocation" over time. From the audiences' perspective, the delay difference between any two users is constantly bounded, and the user experience is enhanced for the entire video.

\section{Numerical Analysis and Results}\label{sect-result}
\subsection{QoE Dataset and Parameter Setting}
Our QoE model in \eqref{Q-i} is synthesized from viewing experience tests in which non-expert volunteers give scores (scaling from 1 to 5) based on their perceived video quality. We follow standardized procedures specified in the Adjectival Categorical Judgment Methods in \cite{recommendation2002500} and the ITU recommendations \cite{itu1999subjective}. The play time of each video is 10s, the same length as a video segment in our setting.
\begin{table}[htbp]
  \centering
  \caption{Mean opinion score (MOS)}
    \begin{tabular}{|r|r|r|r|r|r|}
    \hline
    \multicolumn{2}{|c|}{Duck Video} & \multicolumn{2}{c|}{Crew Video} & \multicolumn{2}{c|}{Ice Video} \bigstrut\\
    \hline
    Rate  & MOS   & Rate  & MOS   & Rate  & MOS \bigstrut\\
    \hline
    19484.0 & 5.0     & 6520.8 & 5.0     & 2302.6 & 5.0 \bigstrut[t]\\
    8108.5 & 5.0     & 2428.8 & 4.0     & 1133.3 & 5.0 \\
    2878.9 & 4.0    & 1275.0  & 3.5   & 573.0   & 4.0 \\
    1311.9 & 3.4   & 622.3 & 3.0     & 336.7 & 3.0 \\
    1177.3 & 3.0     & 466.6 & 3.0     & 270.8 & 3.0 \\
    621.5 & 3.0     & 394.1 & 2.0     & 208.9 & 2.5 \\
    142.2 & 2.0     & 104.5 & 1.4   & 61.4  & 2.0 \\
    117.0 & 2.0     & 72.2  & 1.4   & 42.9  & 2.0 \\
    76.8  & 2.0     & 48.2  & 1.0     & 28.7  & 2.0 \bigstrut[b]\\
    \hline
    \end{tabular}%
  \label{tab:addlabel}%
\end{table}%

In our experiments, three different videos, i.e., duck, crew and ice, representing high, medium and low playback rate respectively are played in H.264/SVC format. Their mean opinion scores (MOS) are shown in Table I. The parameters in \eqref{Q-i} are determined by minimizing the mean squared error (MSE) between the synthetic model and the real data. These parameters and their corresponding MSEs are given in Table II, where $\alpha_3$ is set as one to avoid negative QoE. It can be seen that the logarithmic QoE function is quite accurate for all videos.
\begin{table}[htbp]
  \centering
  \caption{Parameters and accuracy of QoE function}
    \begin{tabular}{crrrrr}
    \hline
          & $\alpha_1$ & $\alpha_2$ & Min Rate & Max Rate & MSE \bigstrut\\
    \hline
    Duck  & 0.634 & 1554.8 & 76.8  & 8108.5 & 0.0514 \bigstrut[t]\\
    Crew  & 0.802 & 419.6 & 48.2  & 2428.8 & 0.057 \\
    Ice   & 0.765 & 297.3 & 28.7  & 1133.3 & 0.161 \bigstrut[b]\\
    \hline
    \end{tabular}%
  \label{tab:addlabel}%
\end{table}%

Our wireless model follows \eqref{rate-i} and we generate independent and identically distributed (i.i.d.) Rayleigh fading channel with normalized average channel gain for each user. Without loss of generality, the channel bandwidth is set as $2$MHz and unit per bit cost is assumed for both the storage and communication. The playback duration of each segment is 10s. To guarantee social experience during watching, the maximum allowed play delay gap $G$ between users is three seconds.
\begin{figure*} \centering
\subfigure[Duck Video (High Rate)]
{ \includegraphics[width=2.2in]{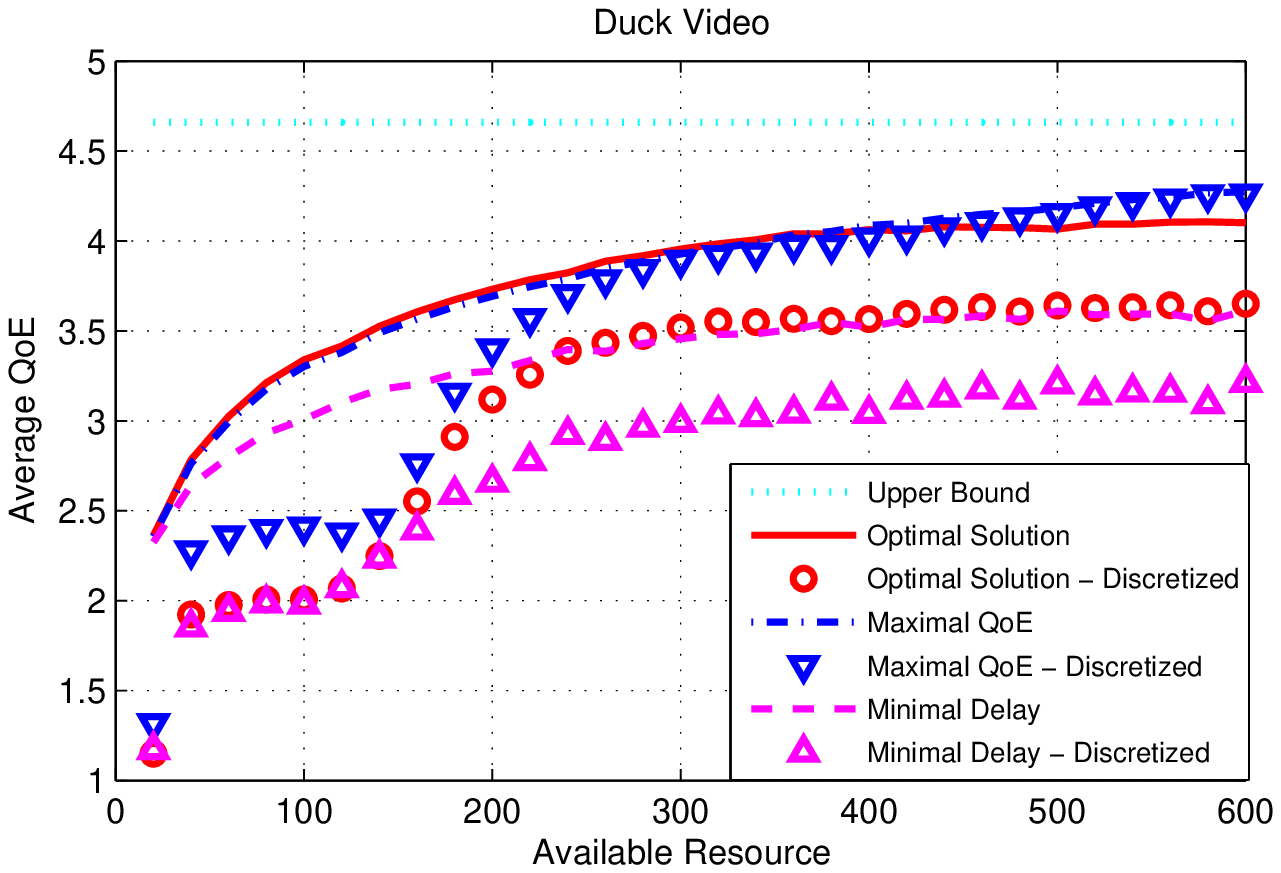} \label{Resource:QoE:Duck} }
\subfigure[Crew Video (Medium Rate)]
{ \includegraphics[width=2.2in]{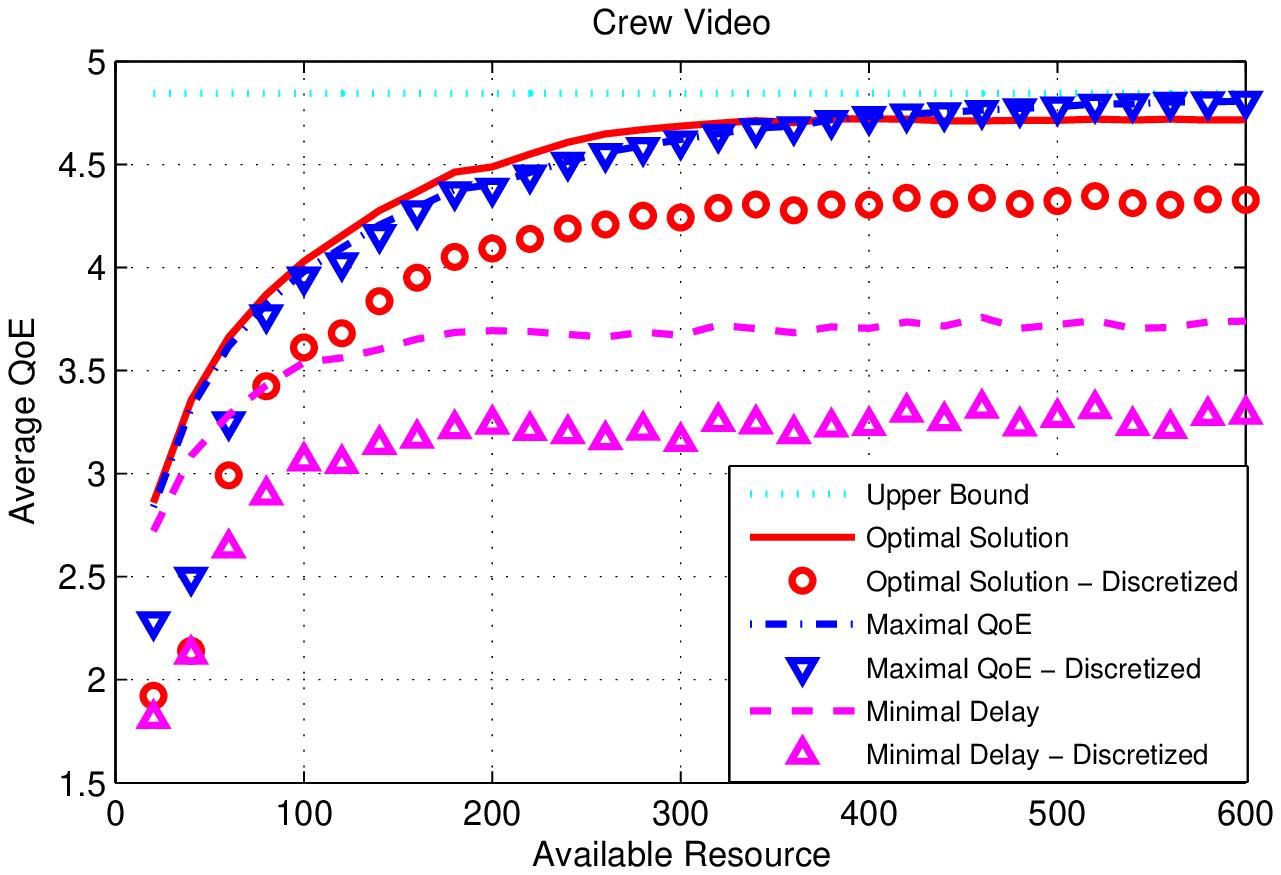} \label{Resource:QoE:Crew} }
\subfigure[Ice Video (Low Rate)]
{ \includegraphics[width=2.2in]{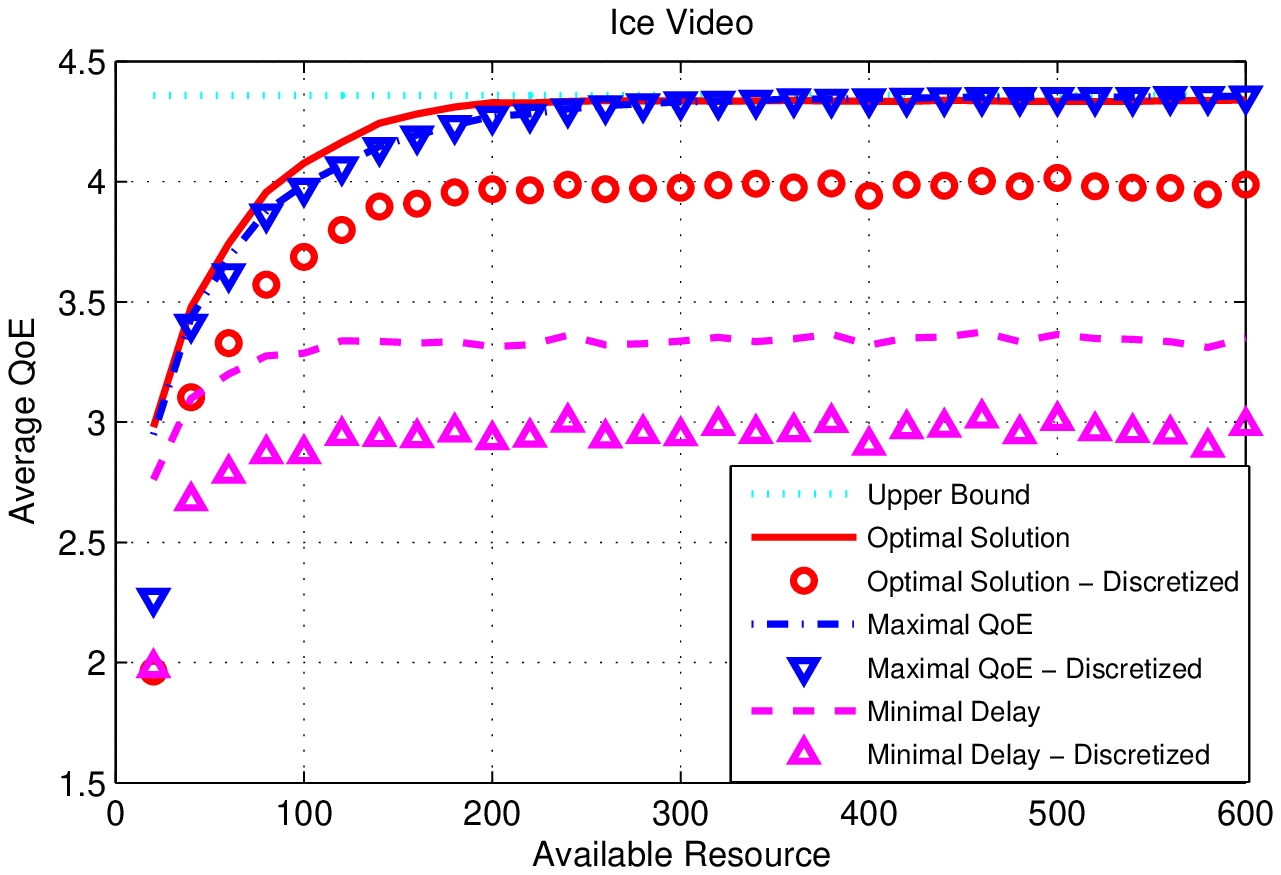} \label{Resource:QoE:Ice} }
\caption{QoE grows as available resource for a Social TV session.} \label{Resource_QoE}
\end{figure*}
\begin{figure*} \centering
\subfigure[Duck Video (High Rate)]
{ \includegraphics[width=2.2in]{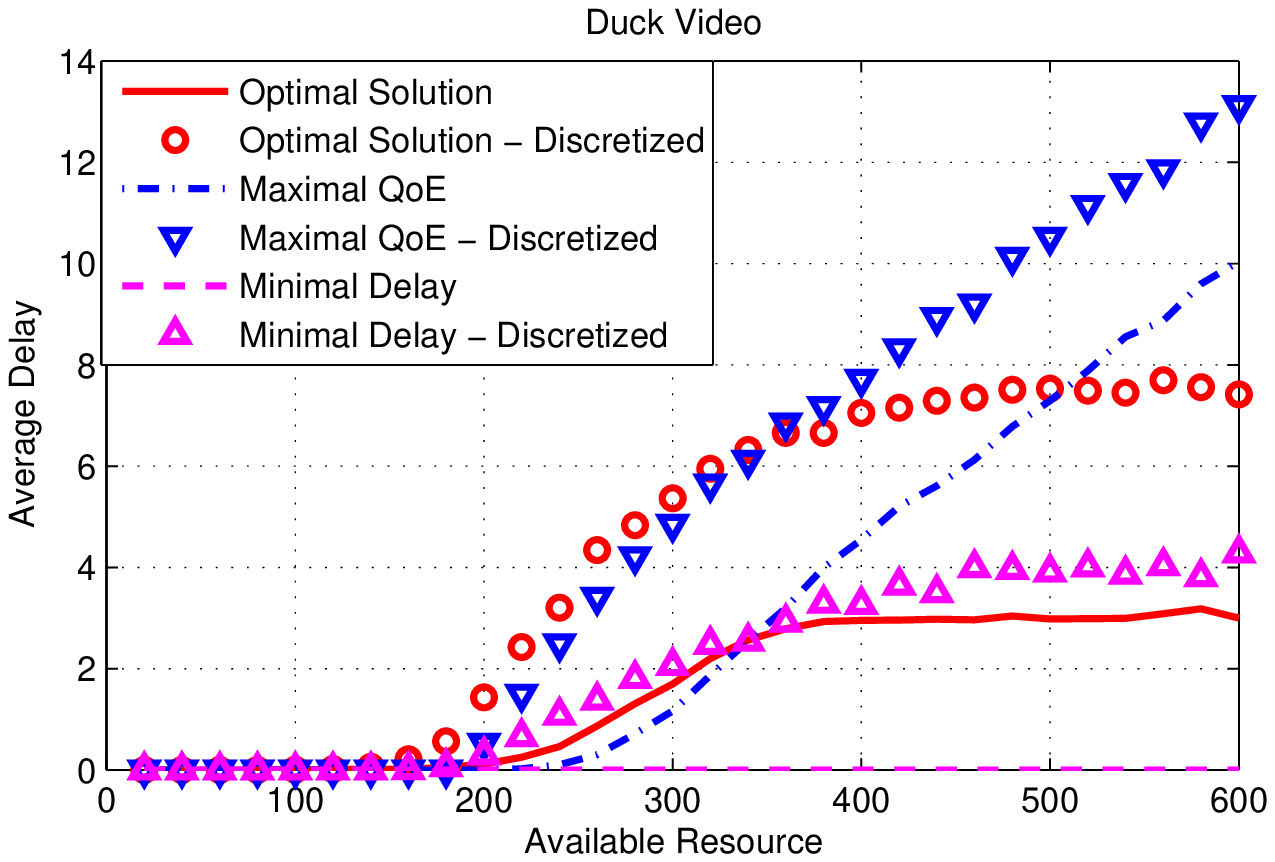} \label{Resource:Delay:Duck} }
\subfigure[Crew Video (Medium Rate)]
{ \includegraphics[width=2.2in]{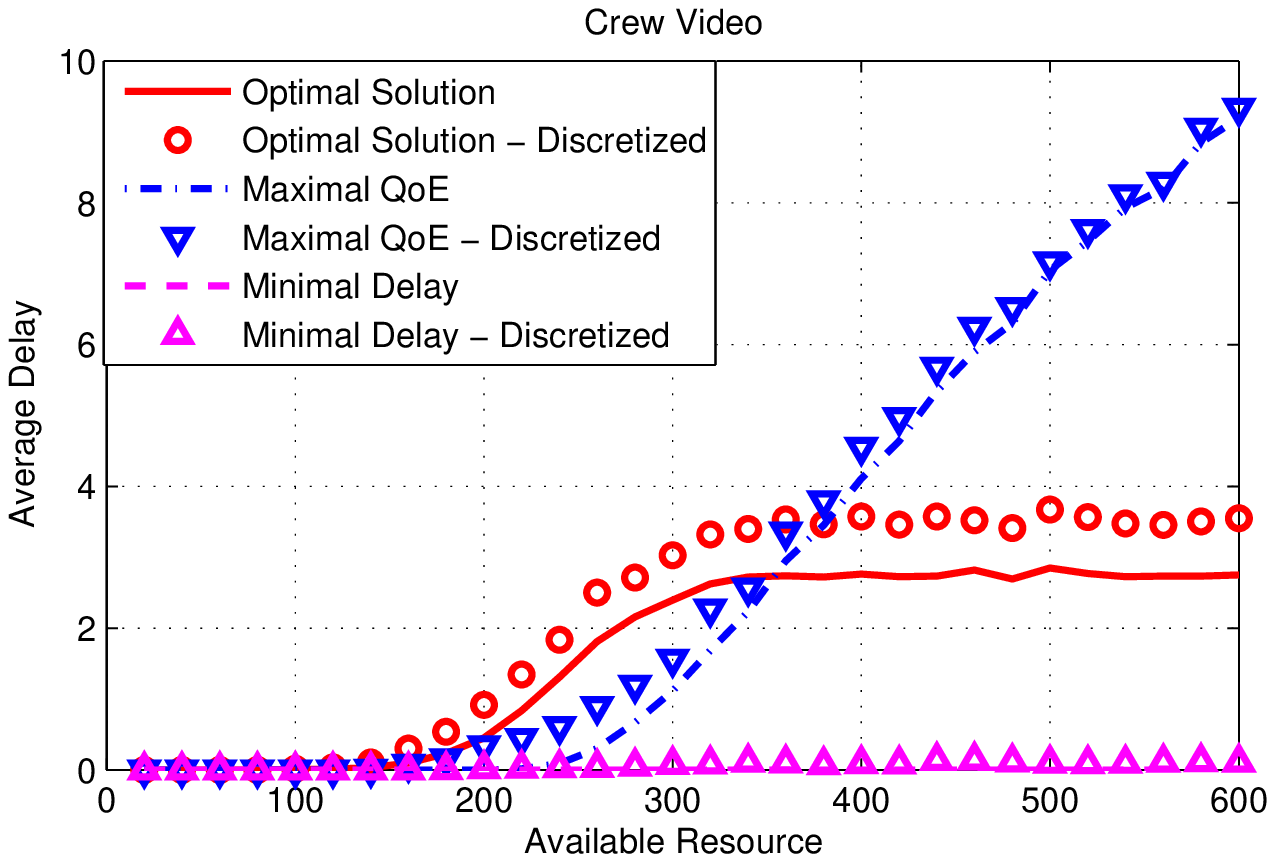} \label{Resource:Delay:Crew} }
\subfigure[Ice Video (Low Rate)]
{ \includegraphics[width=2.2in]{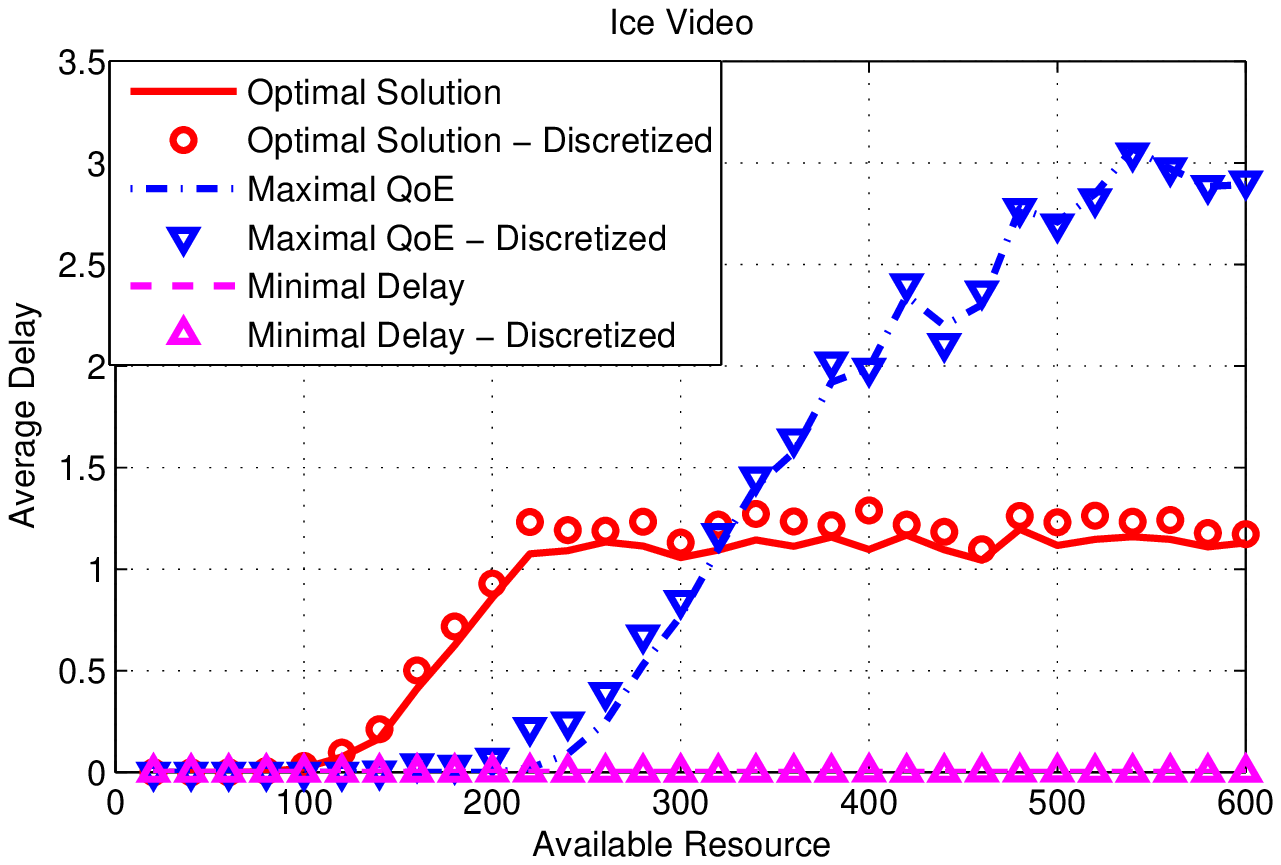} \label{Resource:Delay:Ice} }
\caption{Maximal inter-user delay grows as available resource for a Social TV session.} \label{Resource_Delay}
\end{figure*}

We first briefly describe two extreme rate allocation schemes as performance benchmarks:
\begin{itemize}
  \item \emph{Maximal QoE Scheme:} the Social TV cloud allocates as much available resource as possible to satisfy each user's QoE demand, while completely ignoring the delay constraint. 
  \item \emph{Minimal Delay Scheme:} the Social TV cloud allocates resource proportional to their available transmission rate $R_i$. In this case, each user has the same delay, thus the delay gap is minimized to zero.
\end{itemize}

We first examine the QoE and delay performance versus available resource. Specifically, the available resource is roughly proportional to the maximal playback rate that can be allocated to users. As seen in Fig.~\ref{Resource_QoE}, all resource-QoE curves (solid line, dashed line and dash-dot line) are also roughly governed by the logarithmic rate-QoE function. The maximal QoE scheme and our optimal solution generate the highest QoEs, and the minimal delay scheme has the worst QoE. Fig.~\ref{Resource_Delay} shows the resource-delay curves, in which the minimal delay scheme has zero inter-user delay and the maximal QoE scheme has unbounded inter-user delay. The inter-user delay of our scheme increase with the available resource but is ultimately upper bounded within tolerable value.

Combining Fig.~\ref{Resource_QoE} and Fig.~\ref{Resource_Delay}, it can be seen that the minimal delay scheme forces low inter-user delay at the cost of poor QoE, while the maximal QoE scheme suffers from unbounded delay. For all three video types, our proposed optimal solution achieves near-maximal QoE, while their inter-user delays are strictly controlled within the allowed range, i.e., less than three seconds. Thus, our solution achieves near-optimal tradeoff between higher QoE and lower inter-user delay.

As far as the video type is concerned, the obtained QoE grows faster for videos with lower rate (e.g., the crew and ice videos) and slower for videos with higher rate (e.g., the duck video), and they finally achieve the theoretical upper bound granted sufficient resource. This is intuitive because higher rate videos usually require more resource. Moreover, the QoE gain over the minimal delay scheme is more significant for videos with lower rate, because they have more free resource for optimal rate allocation.

\subsection{Adaptive Rate Allocation}
\begin{figure*} \centering
\subfigure[Duck Video (High Rate)]
{ \includegraphics[width=2.2in]{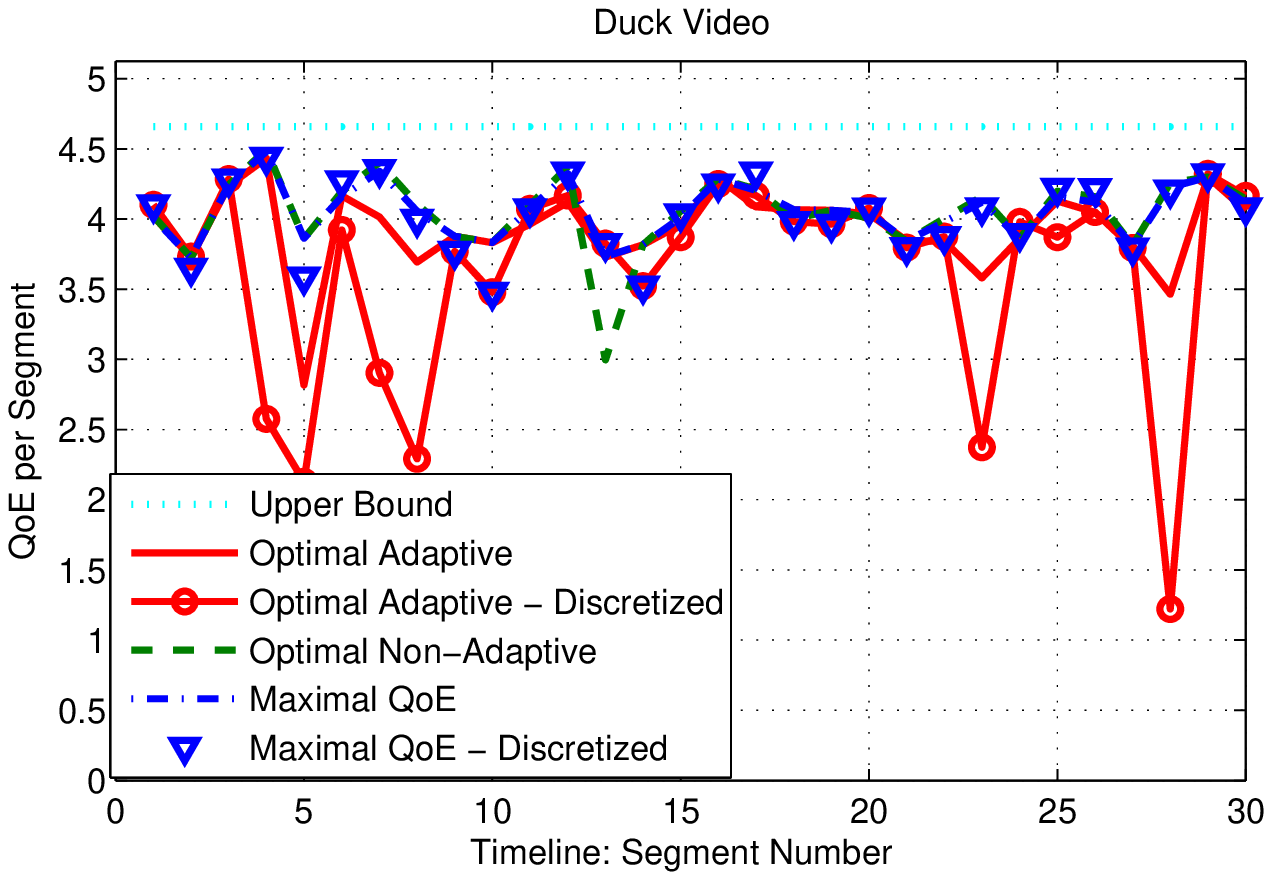} \label{Adaptive:QoE:Duck} }
\subfigure[Crew Video (Medium Rate)]
{ \includegraphics[width=2.2in]{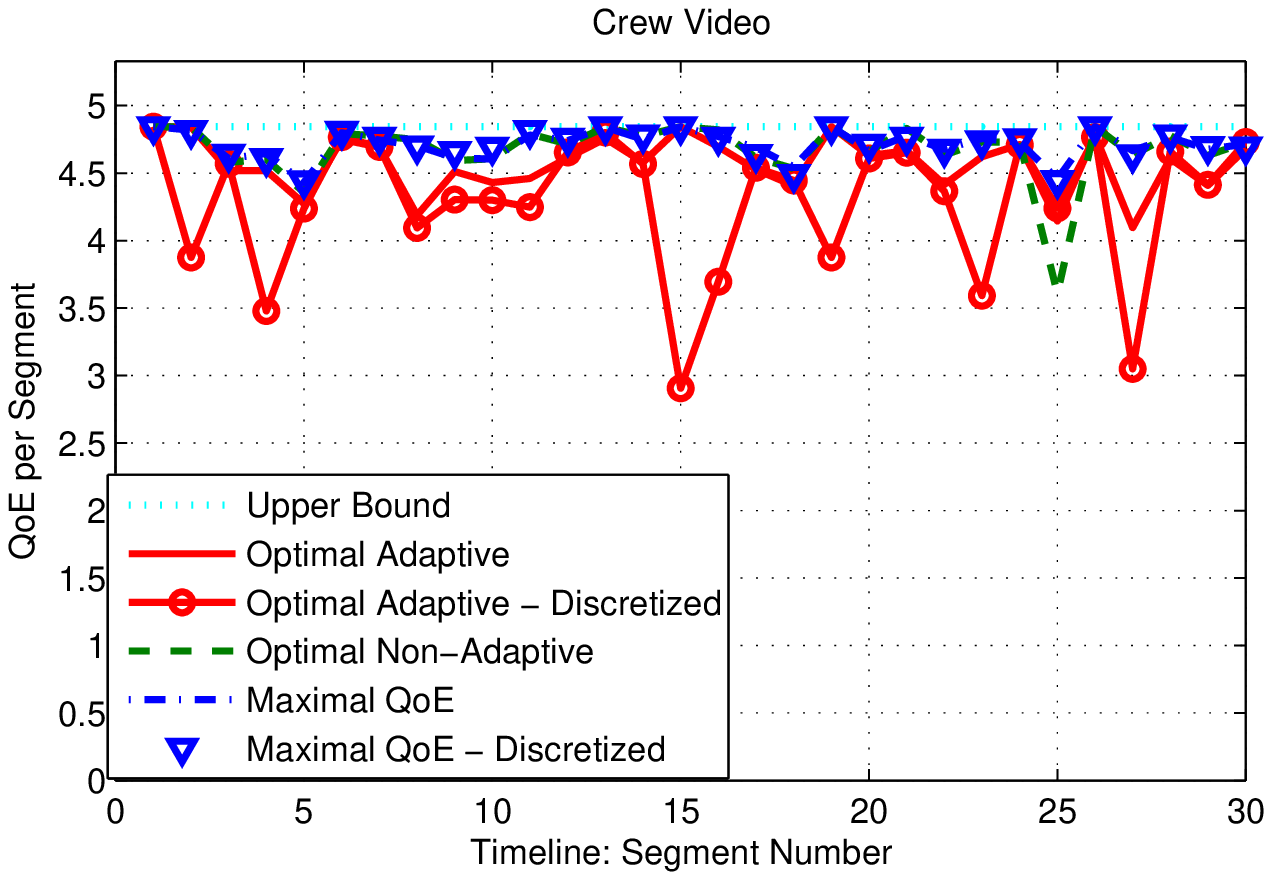} \label{Adaptive:QoE:Crew} }
\subfigure[Ice Video (Low Rate)]
{ \includegraphics[width=2.2in]{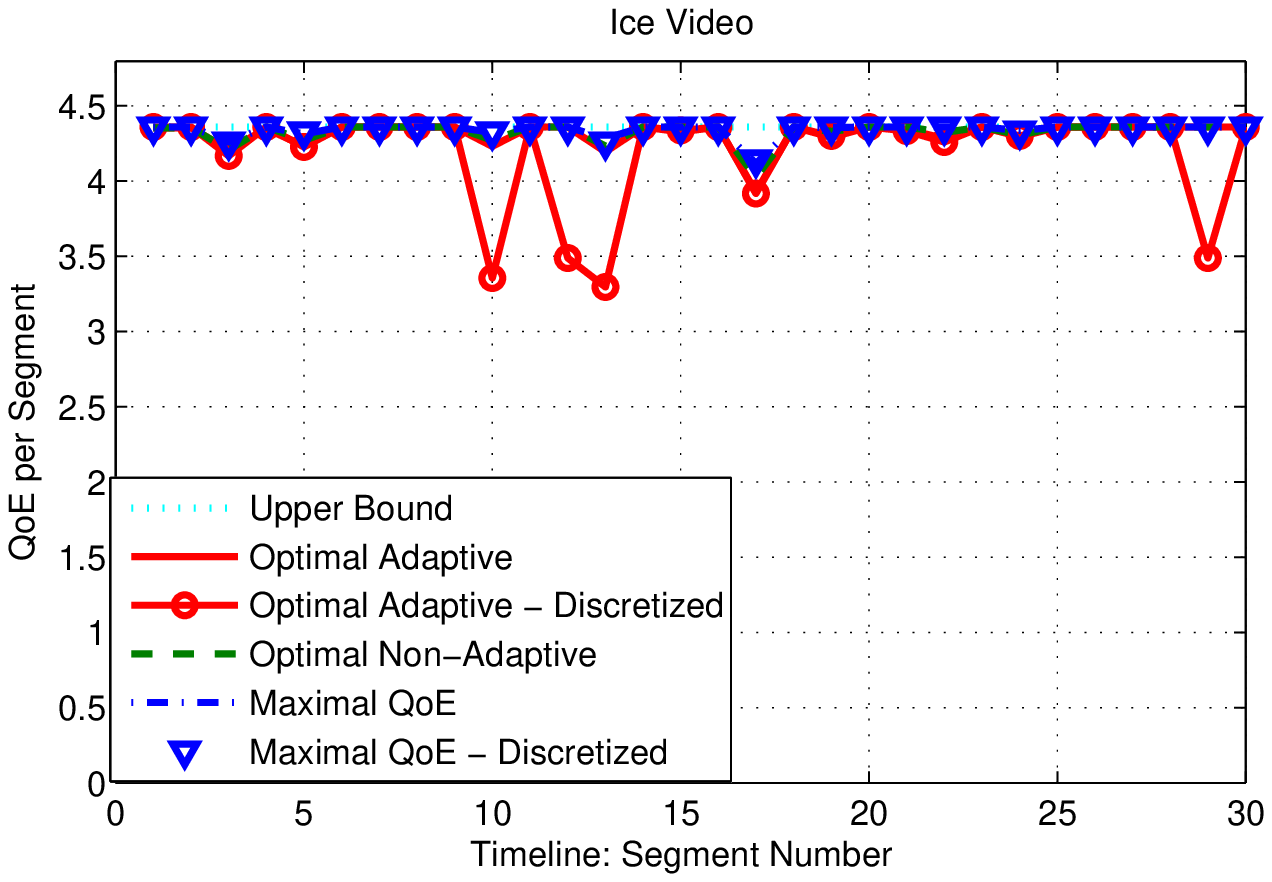} \label{Adaptive:QoE:Ice} }
\caption{Realtime QoE: Adaptive v.s. Non-Adaptive v.s. Maximal QoE.} \label{Adaptive_QoE}
\end{figure*}
\begin{figure*} \centering
\subfigure[Duck Video (High Rate)]
{ \includegraphics[width=2.2in]{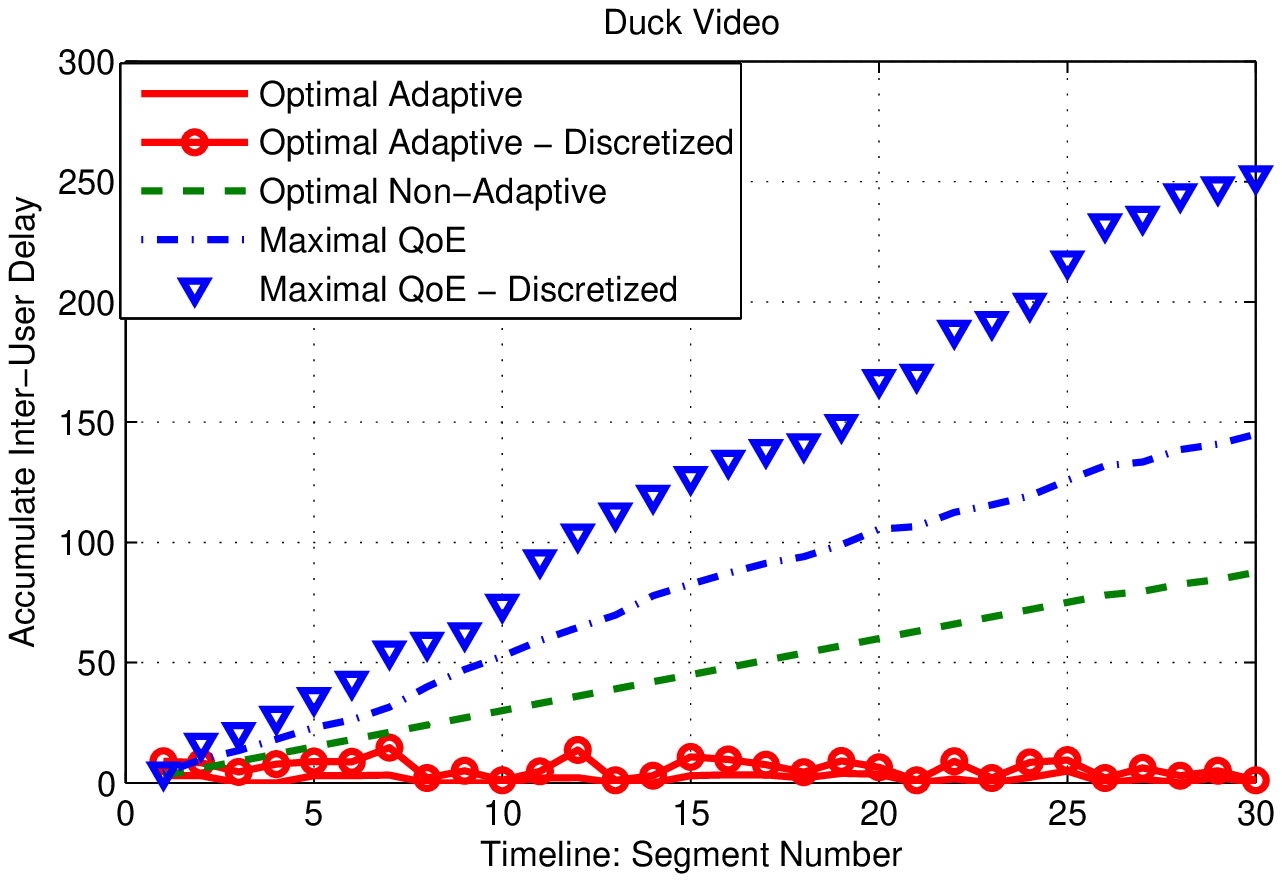} \label{Adaptive:Delay:Duck} }
\subfigure[Crew Video (Medium Rate)]
{ \includegraphics[width=2.2in]{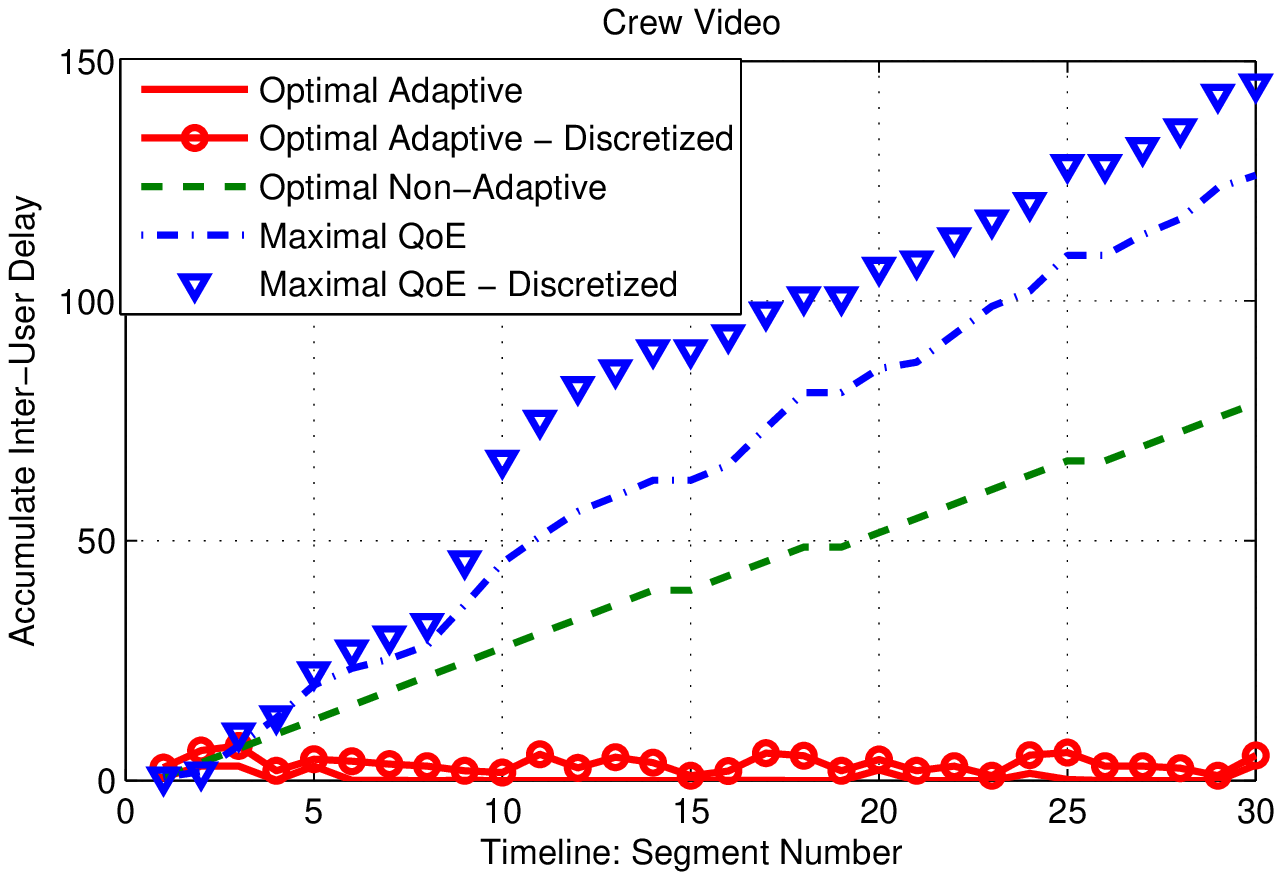} \label{Adaptive:Delay:Crew} }
\subfigure[Ice Video (Low Rate)]
{ \includegraphics[width=2.2in]{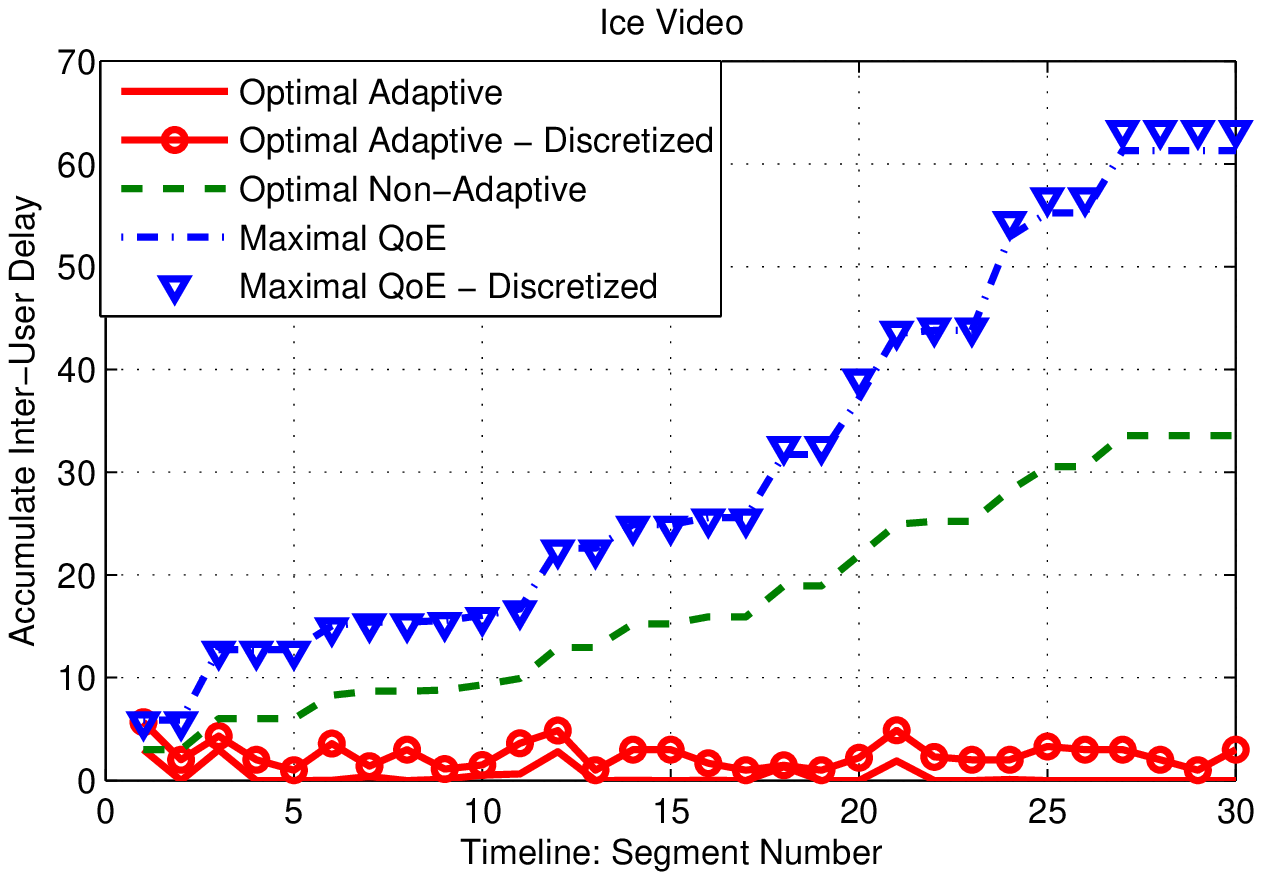} \label{Adaptive:Delay:Ice} }
\caption{Maximal accumulative inter-user delay: Adaptive v.s. Non-Adaptive v.s. Maximal QoE.} \label{Adaptive_Delay}
\end{figure*}
We implement an adaptive rate allocation scheme described at the end of Section \ref{sect-implementation}, by applying the ``accumulative delay constraint" in \eqref{adaptive-constraint}. The essence of the adaptive scheme is that the accumulative inter-user delay, instead of the only the current segment's delay, is taken into account, so that the user experience does not deteriorate over time. The realtime QoE and accumulative inter-user delay are shown in Fig.~\ref{Adaptive_QoE} and Fig.~\ref{Adaptive_Delay}. We generate i.i.d Rayleigh fading for each segment to emulate very fast changing channels. As seen, the adaptive version occasionally incurs additional performance loss from its non-adaptive counterpart, but the overall good performance is preserved. Additionally, the inter-user delay is bounded throughout the video, so that the \emph{virtual living room experience} is also guaranteed.

\subsection{Discrete Video Resolution}
Till now, we assume continuous segment size for both requested and allocated video, so that the potential benefit of the proposed rate allocation scheme may be fully demonstrated. However, our current Social TV framework only support up to five resolutions. Hence, we further simulate the case that the segment size can only take five values. Note that we ensure that the resource constraint is still satisfied when approximating the continuous value to the nearest discrete value. The discretized version of all curve are shown by markers from Fig.~\ref{Resource_QoE} to Fig.~\ref{Adaptive_Delay}. As seen, discretization incurs performance loss in terms of both the QoE, especially for higher rate video. Nevertheless, the good performance is largely preserved, and the discrete scheme is still effective.

\section{Conclusions}\label{sect-conclusion}
In this paper, we introduce the user experience of social TV, and provide an optimal playback rate allocation scheme. Specifically, our scheme can dynamically allocate resource in response to the rapidly changing channel quality and user request. The proposed scheme maximizes the overall QoE while upper bounding inter-user delays among all users, thus ensuring both the perceived video quality and the virtual living room experience. We also provide implementation details, and show through experiment that our algorithm achieves near-optimal tradeoff between video quality and inter-user delay, and performances well even under fast channel fading.

\nocite{*}
\bibliographystyle{IEEEtran}
\bibliography{reference}

\end{document}